\documentclass[preprint,aps,showpacs,]{revtex4}
\usepackage{graphicx}
\begin{document}
\title{Linear and nonlinear properties of Rao-dust-Alfv\'en waves
in magnetized plasmas \footnote{Preprint submitted to
\textit{Physics of Plasmas}. }}
\author{I. Kourakis and P. K. Shukla}
\affiliation{Institut f\"ur Theoretische Physik IV, Fakult\"at
f\"ur Physik und Astronomie, Ruhr--Universit\"at Bochum, D-44780
Bochum, Germany}
\date{28 October 2003}

\begin{abstract}
The linear and nonlinear properties of the
Rao-dust-magnetohydrodynamic (R-D-MHD) waves in a dusty
magnetoplasma are studied. By employing the inertialess electron
equation of motion, inertial ion equation of motion, Amp\`ere's
law, Faraday's law, and the continuity equation in a plasma with
immobile charged dust grains, the linear and nonlinear propagation
of two-dimensional R-D-MHD waves are investigated. In the linear
regime, the existence of immobile dust grains produces the Rao
cutoff frequency, which is proportional to the dust charge density
and the ion gyrofrequency. On the other hand, the dynamics of an
amplitude  modulated R-D-MHD waves is governed by the cubic
nonlinear Schr\"odinger equation. The latter has been derived by
using the reductive perturbation technique and the two-timescale
analysis which accounts for the harmonic generation nonlinearity
in plasmas. The stability of the modulated wave envelope against
non-resonant perturbations is studied. Finally, the possibility of
localized envelope excitations is discussed.
\end{abstract}

\maketitle

\section{Introduction}

A wide variety of electrostatic and electromagnetic oscillatory
modes are known to propagate in unmagnetized and magnetized
plasmas \cite{Krall, Stix}. Since more than a decade ago, it has
been pointed out, and is now well established, that the presence
of heavy charged dust particulates in a plasma may strongly modify
the dispersion properties of the known low-frequency modes, and
may also introduce novel waves \cite{Verheest, PSbook}. For
instance, inclusion of the dust particle dynamics in an
unmagnetized dusty plasma gives rise to the dust-acoustic waves
\cite{Rao}, while the modification of the plasma constituents'
charge balance is responsible for the dust ion-acoustic waves
\cite{SSDIAW}, characterized by an increased phase speed in
comparison with the acoustic speed in an electron-ion plasma
without dust. In a magnetized dusty plasma, a variety of new modes
have been shown to exist, including modified Alfv\'en waves
\cite{PKS1992} propagating along to the direction of the external
magnetic field $\mathbf{B}$, as well as the modified
magnetoacoustic \cite{Rao1993, Rao1995} and drift-electromagnetic
\cite{PKS2003} waves propagating across $\mathbf{B}$.

In this paper, we will focus on the linear and nonlinear
properties of the Rao-dust-magnetohydrodynamic (R-D-MHD) waves
\cite{Rao1995} in two space dimensions. The dispersion
characteristics of the two-dimensional (2D) R-D-MHD waves differ
from the ordinary magnetosonic waves propagating in a magnetized
electron--ion (e--i) plasma; of particular importance is the
existence of a novel cutoff frequency due to the presence of
charged dust grains, as first reported by Rao in his classic paper
\cite{Rao1995}. Apart from being interesting from a fundamental
point of view, and not so widely studied so far, the R-D-MHD waves
have been recently shown \cite{PS2003} to be excited by the
upper-hybrid waves in a uniform dusty magnetoplasma. Our objective
here is twofold: i) to present two-dimensional R-D-MHD modes, ii)
to study the amplitude modulation of finite amplitude 2D R-D-MHD
waves.  Assuming the existence of a uniform external magnetic
field and relying on the two-fluid model description, we will
calculate analytically the harmonic response of the system to a
small displacement from equilibrium, trying to point out the role
of the dust. The nonlinear modulation of the wave's amplitude will
then be considered by making use of an appropriate reductive
perturbation method \cite{redpert, IKPSDIAW, IKPSDAW}.  The
R-D-MHD wave stability will then be investigated and the existence
of envelope excitations will be discussed.

The manuscript is organized in the following fashion. In Sec. II,
we present the governing equations for the R-D-MHD waves.
Linearized equations and harmonic solutions are presented in Sec.
III. Considering oblique nonlinear amplitude modulations of finite
amplitude R-D-MHD waves, we derive the nonlinear Schr\"odinger
equation in Sec. IV.
A stability analysis is carried out in Sec.
V. Section VI contains a discussion of localized R-D-MHD modes.
Our conclusions are highlighted in Sec. VII.

\section{The model}

We consider a three-component fully ionized dusty plasma
composed of electrons (mass $m$, charge $e$), ions (mass $m_i$,
charge $q_i = + Z_i e$) and heavy charged dust particulates (mass $m_d$,
charge $q_d = s\, Z_d e$), henceforth denoted by $e,\, i, \,d$
respectively. Dust mass and charge will be taken to be constant,
for simplicity. Note that both negative and positive dust charge
cases are considered, distinguished by the charge sign $s =$ sgn
$q_d = \pm 1$.

The plasma is immersed in a uniform external magnetic field along
the $\hat z-$direction: $\mathbf{B_0} = B_0\, \hat z$
($B_0 =$ const.)

\subsection{Evolution equations}

Let us consider the MHD system of equations for electrons
and ions. The massive dust particles are assumed to be
practically immobile (`frozen' i.e. $n_d \approx n_{d, 0}$), since
we are interested on timescales much shorter than the dust plasma
period ($\sim \omega_{p, d}^{-1}$). The electron/ion number
density $n_{i, e}$ and velocity $\mathbf{v}_{i, e}$ are governed
by the equations
\begin{equation}
\frac{\partial n_e}{\partial t} + \nabla \cdot (n_e
\,\mathbf{u}_e)= 0 \, ,
\label{el-n-equation}
\end{equation}
\begin{equation}
\frac{\partial n_i}{\partial t} + \nabla \cdot (n_i
\,\mathbf{u}_i)= 0 \, ,
\label{ion-n-equation}
\end{equation}
\begin{equation}
\mathbf{E} +  \frac{1}{c} \mathbf{u_e \times  B} = 0 \, ,
 \label{el-u-equation}
\end{equation}
and
\begin{equation}
m_i \,D_i\, \mathbf{u_i} \, \equiv \, m_i \,\biggl( \frac{\partial
\mathbf{u_i}}{\partial t} + \mathbf{u_i} \cdot \nabla \mathbf{u_i}
\biggr) \, = \, Z_i \,e \,\biggl( \mathbf{E} +  \frac{1}{c}
\mathbf{u_i \times  B} \biggr) \, ,
 \label{ion-u-equation}
\end{equation}
where we have completely ignored the electron inertia, as well as
pressure (temperature) effects (for all species $\alpha$);
the convective derivative operator: $D_i\, \equiv \,
\frac{\partial }{\partial t} + \mathbf{u_i} \cdot \nabla $
has been defined.
 $\mathbf{E}$ and $\mathbf{B}$ denote the (total) electric and
magnetic fields, $\mathbf{E} = \mathbf{0} + \mathbf{E_1}$ and
$\mathbf{B} = \mathbf{B_0} + \mathbf{B_1}$, respectively,
i.e. index 0 (1) denotes the external (wave) field components.
Throughout this text, we shall assume that
$\mathbf{E_1} = (E_{1, x}, E_{1, y}, 0)$ and
$\mathbf{B_1} = (0, 0, B_{1})$, where
$E_{1, x/y}$ and $B_1$ are allowed to depend on $\{ x, y, t \}$.
The system is closed with Maxwell's equations;
neglecting the displacement current, Amp\`ere's law reads
\begin{equation}
\mathbf{\nabla \times B}  =\, \frac{4 \pi}{c} \, \mathbf{J} \equiv
\frac{4 \pi}{c} \, \sum_\alpha q_\alpha\, n_\alpha
\,\mathbf{u_\alpha} = \,\frac{4 \pi e}{c} \,(Z_i \,n_i
\mathbf{u_i} - \,n_e \mathbf{u_e})
 \label{Ampere}
\end{equation}
and Faraday's law is
\begin{equation}
\mathbf{\nabla \times E}  =\, - \frac{1}{c} \, \frac{\partial
\mathbf{B}}{\partial t}\, .
 \label{Faraday}
\end{equation}
Note that the condition $\mathbf{\nabla \cdot B} = 0$ here reduces
to $\partial B/\partial z = 0$. At equilibrium, the overall neutrality
condition holds
\begin{equation}
n_{e,0} - Z_i \, n_{i, 0} \, - s \,Z_d \, n_{d} \, =\, 0 \,.
\label{neutrality}
\end{equation}

Since we are interested in waves propagating in the direction
perpendicular to the magnetic field, will shall assume, throughout
this study, that the velocities $\mathbf{u_\alpha}$ ($\alpha = e,
i$), the wavenumber $\mathbf{k}$ and the electric field
$\mathbf{E}$ lie in the $xy-$plane. See that $\mathbf{E}$ is
orthogonal to $\mathbf{u_e}$ and $\mathbf{B}$, due to
(\ref{el-u-equation}).

\subsection{Reduced system of equations}

By eliminating $\mathbf{E}$ in  (\ref{el-u-equation}),
(\ref{ion-u-equation}), we obtain
\begin{equation}
m_i \,D_i\, \mathbf{u_i} \, = \, Z_i \,\frac{e}{c} \,\bigl(
\mathbf{u_i} - \mathbf{u_e} \bigr)  \mathbf{\times  B}
\label{Diui}
\end{equation}
which, combined with (\ref{Ampere}), in order to eliminate
$\mathbf{u_e}$, i.e.
\begin{equation}
\mathbf{u_e} \, = \, Z_i \,\frac{n_i}{n_e} \mathbf{u_i} \, -
\,\frac{c}{4 \pi e n_e} \, \mathbf{(\nabla \times  B)} \label{ue}
\end{equation}
yields
\begin{eqnarray}
m_i \,D_i\, \mathbf{u_i} \, & = & \, Z_i \,\frac{q_d n_d}{n_e c} \,
(\mathbf{u_i \times  B}) \ + \,\frac{Z_i}{4 \pi n_e} \,
\mathbf{(\nabla \times  B) \times  B} \nonumber \\
& = & \, Z_i \,\frac{q_d n_d}{n_e c} \, (\mathbf{u_i \times  B}) \ +
\,\frac{Z_i}{4 \pi n_e} \, \biggl[ \mathbf{B \cdot \, \nabla B} -
\frac{1}{2} \, \mathbf{\nabla} B^2 \biggr] \,, \label{reduced1}
\end{eqnarray}
where we have used  the quasineutrality condition, $n_{e} - Z_i \,
n_{i} \, - s \,Z_d \, n_{d} \, =\, 0$.  We observe that, to a
first approximation, i.e. assuming very weak magnetic field
non-uniformity, the ions (and the electrons due to (\ref{ue})) are
subjected to a rotation due to the presence of charged dust
grains, as also shown in Ref. \cite{Rao1995, PKS2003}: notice the
Lorentz centripetal force in the right-hand-side of
(\ref{reduced1}), associated with a rotation frequency which is
directly proportional to the dust charge $q_d$ (and vanishes
without it).

Now, by eliminating $\mathbf{E}$ in (\ref{el-u-equation}), (\ref{Faraday})
and using (\ref{ue}), we obtain
\begin{equation}
\frac{\partial\mathbf{B}}{\partial t} \, = \, \mathbf{\nabla
\times} \, \biggl[ \frac{Z_i n_i}{n_e}(\mathbf{u_i \times
B})\biggr] \, - \, \frac{c}{4 \pi e} \, \mathbf{\nabla \times} \,
\biggl[ \frac{1}{n_e}(\mathbf{\nabla \times B}) \mathbf{\times B}
\biggr] \, .\label{reduced2}
\end{equation}
Note that Eqs. (\ref{Diui}) -- (\ref{reduced2})
lead to a novel low-frequency electromagnetic mode, associated
with the presence of charged dust grains, as was recently shown in Ref.
\cite{PKS2003}; cf. Eqs. (4), (6) -- (8) therein.

The system of equations (\ref{reduced1}), (\ref{reduced2}) is not
closed in $\mathbf{B}$ and $\mathbf{u_i}$, since it also involves
$n_e$ and $n_i$ (both variable), unless one limits the analysis to
small (first order) perturbations from equilibrium. Otherwise, for
a consistent description, one should either use the complete
system of Eqs. (\ref{el-n-equation}) -- (\ref{Faraday}) or retain
Eqs. (\ref{el-n-equation}), (\ref{el-u-equation}), (\ref{Ampere}),
(\ref{Faraday}) and (\ref{Diui}) instead. In the following, we
will adopt the former option.

The set of equations (\ref{el-n-equation}) to (\ref{Faraday}) is a
closed system describing the evolution of the state vector
$\mathbf{S} = (n_e, \,n_i, \,\mathbf{u_e}, \, \mathbf{u_i},
\,\mathbf{E}, \,\mathbf{B})$. By assuming that no other vector
quantity has a component along the magnetic field $\mathbf{B} =
B\, \hat z = (B_0 + B_1)\, \hat z$, viz. $\mathbf{E} = \mathbf{0}
+ \mathbf{E_1}\,=\, (E_x, E_y, 0)$, and $\mathbf{u_{e/i}} =
(u_{e/i, x}, u_{e/i, y}, 0)$, \ where $E_{x/y}, u_{x/y}$ and $B_1$
are functions of $\{ x, y; t \}$, we obtain
\begin{equation}
\frac{\partial n_e}{\partial t} + \frac{\partial }{\partial x}
(n_e \,u_{e, x}) + \frac{\partial }{\partial y} (n_e \,u_{e, y}) =
0 \, ,
\label{redeqne}
\end{equation}
\begin{equation}
\frac{\partial n_i}{\partial t} + \frac{\partial }{\partial x}
(n_i \,u_{i, x}) + \frac{\partial }{\partial y} (n_i \,u_{i, y}) =
0 \, ,
\label{redeqni}
\end{equation}
\begin{equation}
E_x = - \frac{1}{c}\, u_{e, y} \,B \, ,
\label{Ex}
\end{equation}
\begin{equation}
E_y = + \frac{1}{c}\, u_{e, x} \,B \, ,
\label{Ey}
\end{equation}
\begin{equation}
m_i \,\biggl( \frac{\partial }{\partial t} + u_{i, x}
\frac{\partial }{\partial x} + u_{i, y} \frac{\partial }{\partial
y} \biggr) u_{i, x} \, = \, Z_i \,e \,\biggl( E_x + \frac{1}{c}
u_{i, y} B \biggr) \, = \, \frac{Z_i e B}{c} \,\bigl( u_{i, y} -
u_{e, y} \bigr) \, ,
\label{redequix}
\end{equation}
\begin{equation}
m_i \,\biggl( \frac{\partial }{\partial t} + u_{i, x}
\frac{\partial }{\partial x} + u_{i, y} \frac{\partial }{\partial
y} \biggr) u_{i, y} \, = \, Z_i \,e \,\biggl( E_y - \frac{1}{c}
u_{i, x} B \biggr) \, = \,-  \frac{Z_i e B}{c} \,\bigl( u_{i, x} -
u_{e, x} \bigr) \, , \label{redequiy}
\end{equation}
\begin{equation}
\frac{\partial B}{\partial y} \, = \, \frac{4 \pi e}{c} \,\bigl(
Z_i n_i u_{i, x} - n_e u_{e, x} \bigr) \, , \label{dBdx}
\end{equation}
\begin{equation}
\frac{\partial B}{\partial x} \, = \, - \frac{4 \pi e}{c} \,\bigl(
Z_i n_i u_{i, y} - n_e u_{e, y} \bigr) \, , \label{dBdy}
\end{equation}
and
\begin{equation}
\frac{\partial E_y}{\partial x} \, - \, \frac{\partial
E_x}{\partial y} \, = - \frac{1}{c} \frac{\partial B}{\partial t}
\, ,\label{last}
\end{equation}
describing the evolution of the 9 scalar quantities: $n_{e}$,
$n_{i}$, $u_{e, x/y}$, $u_{i, x/y}$, $E_{x/y}$ and $B$. Note that
(\ref{Ex}), (\ref{Ey}) can be used to eliminate $\mathbf{E}$ in
(\ref{last}), which then becomes
\begin{equation}
\frac{\partial B}{\partial t} \, = - \frac{\partial (u_{e, x}
\,B)}{\partial x} \, - \, \frac{\partial (u_{e, y} \,B)}{\partial
y} \, .
%\label{last}
\end{equation}
Equations (\ref{redeqne}) -- (\ref{last}) will be the basis of the analysis that
follows.

\section{Linearized equations - harmonic solutions}

By linearizing around the equilibrium state $\mathbf{S_0} = (n_{e,
0}, \,n_{i, 0}, \,\mathbf{0}, \, \mathbf{0}, \,\mathbf{0},
\,\mathbf{B_0})$ viz. $\mathbf{S} = \mathbf{S_0} + \mathbf{S_1}$
and assuming linear perturbations of the form: $\mathbf{S_1} =
\mathbf{\hat S_1}\, \exp i(\mathbf{k x} - \omega t) \, + c. c. \,
= \mathbf{\hat S_1}\, \exp i(k x + k y - \omega t)\, + c. c. \,$
(`$c. c.$' denotes the complex conjuguate) we obtain a new system
of (linear) equations for the perturbation amplitudes ${(\hat
S_1)}_j$. A tedious, yet perfectly straightforward (see in the
Appendix), calculation leads to the equations
\begin{eqnarray}
\omega (i \omega  v_x + \delta \Omega_{c, i}  v_y) &=& i \Omega_{c,
i}^2 L^2 k_x (k_x  v_x + k_y  v_y)
\, ,
\nonumber \\
\omega (i \omega v_y - \delta \Omega_{c, i}  v_x) &=& i \Omega_{c,
i}^2 L^2 k_y (k_x v_x + k_y  v_y) \, ,
\label{vxvy}
\end{eqnarray}
in terms of the ion velocity component amplitudes
$v_j = \hat u_{i1, j}$ ($j = x, y$), where we have
defined

- the ion gyrofrequency: $\Omega_{c, i} = \frac{Z_i e B_0}{m
c }$ ,

- the characteristic length: $L = \biggl( \frac{m_i c^2 n_{i,
0}}{4 \pi e^2 n_{e, 0}^2} \biggr)^{1/2}$ , and

- the (dimensionless) dust parameter: \( \delta = \frac{Z_{d}
n_{d, 0}}{n_{e,0}} = s \biggl( 1 - \frac{Z_{i} n_{i,
0}}{n_{e,0}}\biggr) \); see that $\delta$
cancels in the dust-free limit [cf. (\ref{neutrality})].

Equations (\ref{vxvy}a, b) constitute a $2 \times 2$ homogeneous
Cramer (linear) system, in terms of $u_x$, $u_y$,
whose determinant should vanish in order for a non-trivial solution
 to exist; the wave frequency $\Omega$ and wavenumber $\mathbf{k}$ are
thus found to obey the dispersion relation
\begin{equation}
\omega^2 =  \omega_g^2 + C^2 k^2
\label{dispersion}
\end{equation}
where $k = (k_x^2 + k_y^2)^{1/2}$;
we have defined
\newline
- the `gap frequency' $\omega_g$
\begin{equation}
\omega_g = \frac{Z_{d} n_{d, 0} Z_i e B_0}{n_{e,0} m_i c} = \delta
\,\Omega_{c, i}  \label{gap}
\end{equation}
and \newline
- the characteristic velocity $C = \Omega_{c, i} \,L$, given
by
\begin{equation}
C^2 = \frac{Z_i^2 B_0^2 n_{i, 0}}{4 \pi n_{e,0}^2 m_i} =
\biggl(\frac{Z_i n_{i, 0}}{n_{e,0}}\biggr)^2 \frac{B_0^2}{4 \pi
n_{i,0} m_i} \equiv (1 - s \delta)^2 \,V_A^2 \, ,
\label{char-disp-velocity}
\end{equation}
i.e. $C \, = \, \Omega_{c, i} \, L \equiv (1 - s \delta) \, V_A$,
where $V_A = B_0/(4 \pi n_{i,0} m_i)^{1/2}$ is the Alfv\'en
speed. Notice the effect of the dust, which results in

- a finite (\emph{`gap'}) oscillation frequency at the infinite
wavelength ($k \rightarrow 0$) limit, and

- a modified phase speed $v_{ph} = \omega/k \, \, \, \, (\ne
v_g = C^2 k/\omega$, \ for $\delta \ne 0)$; as a matter of fact,
the phase speed $v_g$ ($\approx C$ for
$\omega \gg \omega_g$) is higher (lower) than the Alfv\'en
speed $V_A$ in the presence of negative (positive) dust.

Notice that (\ref{dispersion}) coincides with (9) in Ref.
\cite{PKS2003}. It should also be pointed out that the existence of both
the cutoff frequency $\omega_g$ and the modified Alfv\'en speed $C$,
associated with the dust-magnetosonic waves, was predicted
for the first time by Rao in his classic paper \cite{Rao1995}.

The harmonic perturbation amplitudes $\hat S_{1, j}$ may now be
calculated. Assuming $\mathbf{k} = (k \cos\theta, \,k
\sin\theta)$, one obtains the following relations
\begin{equation}
\hat n_{e} \, = \, \frac{n_{e, 0}} {B_0}   \,\hat B_{1} \equiv
c^{(11)}_1 \,\hat B_{1}
 \, ,
\label{ne1-solution}
\end{equation}
\begin{equation}
\hat n_{i} \, = \, \frac{n_{e, 0}} {Z_i B_0} \,\hat B_{1}  \equiv
c^{(11)}_2 \,\hat B_{1}
 \, ,
\label{ni1-solution}
\end{equation}
\begin{equation}
\hat u_{e, x} \, = \, \biggl\{ \omega  \cos\theta - \, i \,
\Omega_{c, i}^{-1} \,\biggl[ \frac{n_{e, 0}} {Z_i n_{i, 0}}
 (\omega^2 - \Omega_{c, i}^2) +
 \Omega_{c,
i}^2 \biggr] \sin\theta \biggr\}
 \, \frac{\hat B_{1}}{B_0 k} \equiv c^{(11)}_3\,\hat B_{1}
 \, ,
\label{uex1-solution}
\end{equation}
\begin{equation}
\hat u_{e, y} \, = \, \biggl\{ \omega  \sin\theta + \, i \,
\Omega_{c, i}^{-1} \,\biggl[ \frac{n_{e, 0}} {Z_i n_{i, 0}}
 (\omega^2 - \Omega_{c, i}^2) +
 \Omega_{c,
i}^2 \biggr] \cos\theta \biggr\}
 \, \frac{\hat B_{1}}{B_0 k} \equiv c^{(11)}_4\,\hat B_{1}
 \, ,
\label{uey1-solution}
\end{equation}
\begin{equation}
\hat u_{i, x} \, = \, \frac{n_{e, 0}} {Z_i n_{i, 0}} \, \bigl[
\omega \, \cos\theta + \, i\, s \delta \Omega_{c, i} \sin\theta
\bigr] \,\frac{\hat B_{1}}{B_0 k} \equiv c^{(11)}_5 \,\hat B_{1}
 \, ,
\label{uix1-solution}
\end{equation}
\begin{equation}
\hat u_{i, y} \, = \, \frac{n_{e, 0}} {Z_i n_{i, 0}} \, \bigl[
\omega \, \sin\theta - \, i\, s \delta \Omega_{c, i} \cos\theta
\bigr] \,\frac{\hat B_{1}}{B_0 k} \equiv c^{(11)}_6 \,\hat B_{1}
 \, ,
\label{uiy1-solution}
\end{equation}
\begin{equation}
\hat E_{x} \, = - \frac{B_0}{c} u_{e, y} = \, - \biggl\{ i \,
\Omega_{c, i}^{-1}\, \biggl[ \frac{n_{e, 0}}{Z_i n_{i,
0}}\,(\omega^2 - \Omega_{c, i}^2) + \Omega_{c, i}^2 \biggr]\,
\cos\theta + \, \omega \, \sin\theta \biggr\} \,\frac{\hat B_{1}}
{c\,k} \, \equiv c^{(11)}_7 \,\hat B_{1} \label{Ex1-solution}
\end{equation}
and
\begin{equation}
\hat E_{y} \, = \frac{B_0}{c} u_{e, x} = \, - \biggl\{ i \,
\Omega_{c, i}^{-1}\,  \biggl[ \frac{n_{e, 0}}{Z_i n_{i,
0}}\,(\omega^2 - \Omega_{c, i}^2) + \Omega_{c, i}^2 \biggr]\,
\sin\theta - \, \omega \, \cos\theta \biggr\} \,\frac{\hat B_{1}}
{c\,k} \,  \equiv c^{(11)}_8 \,\hat B_{1} \label{Ey1-solution} \,
\end{equation}
 (obviously, $c_9^{(11)} = 1$). Note that these relations satisfy
\[
\mathbf{\hat E_{1} \cdot \hat u_{e, 1}} \, = \, 0 \, ,
\]
in agreement with (\ref{el-u-equation}); also,
\[
\mathbf{k \cdot \hat u_{e, 1}} \, = \, \frac{Z_i n_{i, 0}}{n_{e,
0}}\, \mathbf{k \cdot \hat u_{i, 1}} \, = \,\frac{\omega}{B_0} \,
\hat B_1 \,
\] as expected (see the Appendix) as well as
\[
\hat u_{i, 1} = \biggl( 1 - \frac{\omega^2}{\Omega_{c, i}^2}
\biggr)^{-1/2} \,\hat u_{e, 1} \, ,
\]
(remember that the amplitudes $u_{e/i, \,x/y}$ are complex)
implying that the ions
and electrons oscillate in (out of) phase for $\omega$ lower
(higher) than $\Omega_{c, i}$, i.e. for wavenumber values $k$
below (above) a threshold $k_{cr} = Z_i^{1/2} \frac{\Omega_{p,
i}}{c} \bigl( \frac{1 + s \delta}{1 - s \delta} \bigr)^{1/2}$ (see
that $k_{cr} \rightarrow 0$ in the case of complete electron
depletion in the plasma, i.e. $\delta \rightarrow 1$, $s = -1$).

\section{Oblique nonlinear amplitude modulation}

Let us consider the system (\ref{redeqne}) -- (\ref{last}),
which describes the evolution of the (nine scalar)
components of $\mathbf{S}$: $\{ n_e, \,n_i ; \,u_{e, x}, \, u_{e,
y} ; \, u_{i, x}, \,u_{i, y} ; \, \,E_{x}, \, E_{y} ; \, B_1 \}$.

In order to study the amplitude modulation of the R-D-MHD waves
presented in the previous section, we will assume small deviations
from the equilibrium state $\mathbf{S}^{(0)} = \{ n_{e, 0}, \,n_{i, 0} ;
\,0, \, 0 ; \, 0, \,0 ; \, 0, \, 0 ; \, B_0 \}$ by taking
\[
\mathbf{S} = \mathbf{S}^{(0)} \, +
\, \epsilon \, \mathbf{S}^{(1)} +
\, \epsilon^2 \, \mathbf{S}^{(2)}
+ \, ... = \mathbf{S}^{(0)} \, + \,
\, \sum_{n=1}^\infty \epsilon^n \,
\mathbf{S}^{(n)} \, ,
\]
where $\epsilon \ll 1$ is a smallness parameter. Following the
standard multiple scale (reductive perturbation) technique
\cite{redpert, IKPSDIAW}, we shall consider the stretched (slow) space and
time variables
\begin{equation}
\zeta \,= \, \epsilon (x - \lambda \,t) \, , \qquad \tau \,= \,
\epsilon^2 \, t\, , \label{slowvar}
\end{equation}
where $\lambda$, having dimensions of velocity, is a real
parameter to be later defined. In order to allow for an oblique
amplitude modulation on the R-D-MHD wave, we will assume that all
perturbed states depend on the fast scales via the phase $\theta_1
= \mathbf{k \cdot r} - \omega t = k_x x + k_y y - \omega t$ only,
while the slow scales enter the argument of the $l-$th harmonic
amplitude $S_l^{(n)}$, allowed to vary only along $x$,
\[
\mathbf{S}^{(n)} \,= \, \sum_{l=-\infty}^\infty \,
\mathbf{S}_l^{(n)}(\zeta, \, \tau)
 \, e^{i l (\mathbf{k \cdot r} - \omega t)} \, .
\]
%\cite{Kako};
The reality condition
$\mathbf{S}_{-l}^{(n)} = {\mathbf{S}_l^{(n)}}^*$
is met by all state variables. Note that the (choice of) direction of the
propagation remains arbitrary, yet modulation is allowed to take
place in an oblique direction, characterized by the angle variable
$\theta$. Accordingly, the wave-number vector $\mathbf{k}$ is
taken to be \( \mathbf{k} = (k_x, \, k_y) = (k\, \cos\theta, \,
k\, \sin\theta) \). According to these considerations, the
derivative operators in the above equations are treated as follows
\[
\frac{\partial}{\partial t} \rightarrow \frac{\partial}{\partial
t} - \epsilon \, \lambda \, \frac{\partial}{\partial \zeta} +
\epsilon^2 \, \frac{\partial}{\partial \tau} \, ,
\]
and
\[
\mathbf{\nabla} \rightarrow \mathbf{\nabla} + \epsilon \, \hat x
\, \frac{\partial}{\partial \zeta} \equiv (\nabla_x + \epsilon \,
\frac{\partial}{\partial \zeta}, \, \nabla_y)\, ,
\]
i.e. explicitly
\[
\frac{\partial }{\partial t} \, A_l^{(n)} \, e^{i l \theta_1} =
\biggr( - i l \omega \, A_l^{(n)} \, - \epsilon \, \lambda \,
\frac{\partial A_l^{(n)}}{\partial \zeta} + \epsilon^2 \,
\frac{\partial A_l^{(n)} }{\partial \tau} \biggr) \, e^{i l
\theta_1} \, ,
\]
\[
\nabla_x \, A_l^{(n)} \, e^{i l \theta_1} = \bigr( i l k
\cos\theta \, A_l^{(n)} \, + \epsilon \, \hat x \, \frac{\partial
A_l^{(n)}}{\partial \zeta}  \bigr) \, e^{i l \theta_1}  \, ,
\]
and
\[
\nabla_y \, A_l^{(n)} \, e^{i l \theta_1} =  i l k \sin\theta \,
A_l^{(n)}  \, e^{i l \theta_1}  \, ,
\]
for any of the components $A_{l, j}^{(n)}$ ($j = 1, ..., 9$) of
$\mathbf{S}_l^{(n)}$.

By substituting the above expressions into Eqs. (\ref{redeqne}) --
(\ref{last}) and isolating distinct orders in $\epsilon$, we
obtain a set of (nine) reduced equations at each ($n$th-) order,
describing the evolution of the (nine) components of
$\mathbf{S}^{(n)}$. The system is then solved (for each harmonic
$l$), substituted into the subsequent order, and so forth.  This
is a rather standard procedure in the reductive perturbation
method framework \cite{redpert, IKPSDIAW, IKPSDAW}, and we shall
not burden the presentation with unnecessary details. The outcome
of the long algebraic calculation is presented in the following,
while essential details are presented in the Appendix.

The first order ($n = 1$) first harmonic ($l = 1$) equations are
just as described in the previous section. Recall the (parabolic)
form of the dispersion relation (\ref{dispersion}), which arises
as a compatibility condition. The amplitudes of the first
harmonics of the perturbation, say $A_{1, j}^{(1)}$ ($j = 1, ...,
9$) (i.e. precisely $\hat A_{j, 1}$ in the previous Section), then
come out to be directly proportional to the magnetic field
perturbation, viz. $A_{1, j}^{(1)} = c_{j}^{(11)} B_1^{(1)}$; the
coefficients $c_{j}^{(11)}$ are defined in (\ref{ne1-solution}) --
(\ref{Ey1-solution}) above. Only the first harmonics have a
contribution at this order; indeed, for $n = 1, l=0$, one obtains
a ($6 \times 6$) linear homogeneous system of equations for the (6
components of) $\mathbf{u_e}, \mathbf{u_i}, \mathbf{E}$;
interestingly, the determinant $D_0^{(1)} \sim \Omega_{c, i}
q_{d}^2$ is non-zero due to (and only in) the presence of dust, so
we obtain the trivial solution for the zeroth-harmonic
contribution, $\mathbf{u}_{e, 0}^{(1)} = \mathbf{u}_{i, 0}^{(1)} =
\mathbf{E}_0^{(1)} = \mathbf{0}$. In addition, $n_{e, 0}^{(1)} =
n_{i, 0}^{(1)} = B_{0}^{(1)} = 0$, as imposed by the ($n = 2,
l=0$) equations.

\subsection{Second order in $\epsilon$:
group velocity, 0th and 2nd harmonics}

The second order ($n = 2$) equations for the first harmonics
provide the compatibility condition: $ \lambda \,  =
{\partial \omega}/{\partial k_x} = \omega'(k) \cos\theta$,
which defines $\lambda$ as the
group velocity $v_g = (C^2 k/\omega) \,
\cos\theta$ (the characteristic velocity $C$ was defined previously).
The 2nd-order corrections to the first
harmonic amplitudes, say
$A_{1, j}^{(2)}$ ($j = 1, ..., 9$), come out to be
$ A_{1, j}^{(2)} = c_{j}^{(21)} {\partial B_1^{(1)}}/{\partial
\zeta}$, where the coefficients $c_{j}^{(21)}$ are presented
in the Appendix.

As expected, second order harmonic contributions arise in this
order; their  amplitudes, defined by the equations for $n = 2$, $l
= 2$, are found to be proportional to the square of the first
order elements, e.g. in terms of $B_1^{(1)}$: $A_{2, j}^{(2)} =
c_{j}^{(22)} ({B_1^{(1)}})^2$. The nonlinear self-interaction of
the carrier wave also results in the creation of a zeroth
harmonic, to this order; its strength is analytically determined
by taking into account the $l = 0$ component of the 3rd and 4th
order reduced equations. The result is conveniently expressed in
terms of the square modulus of the ($n = 1$, $l = 1$) quantities,
e.g. in terms of $|B_1^{(1)}|^2  = (B_1^{(1)})^*\, B_1^{(1)}$,
viz. $A_{0, j}^{(2)} = c_{j}^{(22)} \,|B_1^{(1)}|^2$ ($j = 1, ...,
9$); once more, the definitions of $c_{j}^{(22)}$, $c_{j}^{(20)}$
can be found in the Appendix.  Notice (see the Appendix) the
dependence of the expressions derived in this Section (except
those for $n_{e, i}$, $B$, in fact) on the value of $\theta$.

\subsection{Derivation of the Nonlinear Schr\"odinger Equation}

Proceeding to the third order in $\epsilon$ ($n=3$), the equation
for $l = 1$ yields an explicit compatibility condition to be
imposed in the right-hand side of the evolution equations which,
given the expressions derived previously, can be cast into the
form of the Nonlinear Schr\"odinger Equation (NLSE)
\begin{equation}
i\, \frac{\partial \psi}{\partial\tau} + P\,
\frac{\partial^2 \psi}{\partial\zeta^2} + Q \, |\psi|^2\,\psi = 0
\, .
\label{NLSE}
\end{equation}
where $\psi\, \equiv \,  B_1^{(1)}$ denotes the amplitude of the
first-order electric field perturbation. Recall that the
`slow' variables $\{ \zeta, \tau \}$ were defined in
(\ref{slowvar}).

The {\em dispersion coefficient} $P$ is related to the curvature
of the dispersion curve as \( P \,  = \, \frac{1}{2} \,
\frac{\partial^2 \omega}{\partial k_x^2} \,= \, \frac{1}{2}\,
\biggl[ \omega''(k) \, \cos^2\theta \, + \omega'(k) \,
\frac{\sin^2\theta}{k} \biggr] \); the exact form of P reads
\begin{equation}
P(k) \,  =\, \frac{C^2}{2 \omega^3} \, \biggl(\omega_g^2 \,
\cos^2\theta \, + \, \omega^2 \, \sin^2\theta \biggr) \, ,
\label{Pcoeff}
\end{equation}
which is positive for all values of the angle $\theta$, as
expected from the parabolic form of $\omega(\mathbf{k})$.

The {\em nonlinearity coefficient} $Q$ is due to
the carrier wave self-interaction. It is given by
\begin{eqnarray}
Q = \, \frac{\omega}{4 \,B_0^2 \,c^2 \,e^2 \,m_i \,n_{i, 0}^2 \, \pi \,Z_i^2
(n_{e, 0} - Z_i n_{i, 0})^2 \, k^2}   \,
\biggl[
- 3 m_i^2 c^4 \, n_{e, 0}^2 n_{i, 0} \, k^4 \, \qquad \qquad \qquad
\qquad \qquad
\nonumber \\
-
32 e^4 \, \pi^2 \, Z_i\, (n_{e, 0} - Z_i n_{i, 0})^4\,
(n_{e, 0} + Z_i n_{i, 0}) \, +
4 c^2 e^2 \, k^2 \, m_i \, \pi \, (n_{e, 0} - Z_i n_{i, 0})^2
(n_{e, 0}^2 - Z_i^2 n_{i, 0}^2)
\biggl]
\, .
\label{Qcoeff}
\end{eqnarray}
Quite surprisingly, $Q$ comes out to be independent of the angle $\theta$.
However, as expected, the presence of charged dust grains in the charge balance
equation (\ref{neutrality}) strongly affects the numerical value of $Q$; notice, in
passing, that this expression is not valid in the absence of dust grains
(since the denominator then vanishes).

The last expression for $Q$ can be conveniently re-arranged, by
making use of appropriate plasma quantities. Let us first define
the dust parameter: $\mu = n_{e,0}/(Z_i n_{i,0})$; see that: \(
\mu \, = \, 1 + s\, ({Z_{d} n_{d,0}})/({Z_{i} n_{i,0}})\), due to
(\ref{neutrality}), so a value lower/higher than $1$ corresponds
to negative/positive dust charge sign; $\mu$ obviously tends to
unity in the absence of dust (in any case, $\mu \ge 0$). Check
that $\mu = (1 - s \delta)^{-1}$ [or $\delta = s (1 - 1/\mu)$],
where $\delta$ was defined above. By normalizing the wavenumber
$k$ as $k = K \, \omega_{p, i}/c \equiv (4 \pi n_{i, 0} Z_i
e^2/m_i c^2)^{1/2} K$ ($\omega_{p, i}$ is the ion plasma
frequency), expression (\ref{Qcoeff}) can be cast into an elegant
form
\begin{equation}
Q(K, \mu) = \, \frac{[(\mu - 1)^2 + x^2]^{1/2} \, Z_i^2 e^2} {\mu
\,(\mu - 1)^2 \,m_i^2 \,c^2 \, \Omega_{c, i} \,K^2}   \, \biggl[-
3 \mu^2 K^4 \, + (\mu - 1)^2 (\mu^2 - 2) \,K^2 - 2 (\mu - 1)^4
(\mu + 1)
 \biggr]
\,
\label{Qreduced}
\end{equation}
($\Omega_{c, i}$ denotes the ion gyrofrequency defined previously).
Retaining the approximate long-wavelength (i.e. vanishing wavenumber) behaviour of
$Q$, we have
\begin{equation}
Q(K \ll 1, \mu) \approx \, - 2
\frac{(1-\mu)^3 (1 + \mu) \,Z_i^2}
{\mu \,m_i^2 \,c^2 \, \Omega_{c, i} \,K^2}
\,
\label{Qapprox}
\end{equation}
which is always negative and thus ensures, as we shall see in the
following, stability at long wavelengths. Note, for later
reference, that the same scaling results in relations
(\ref{dispersion}) and (\ref{Pcoeff}) taking, respectively, the
reduced forms
\begin{equation}
\omega = \Omega_{c, i}\, \biggl[ \biggl( 1 - \frac{1}{\mu}
\biggr)^2 + \frac{x^2}{\mu} \biggr]^{1/2} \label{omegareduced}
\end{equation}
and
\begin{equation}
P = \frac{c^2 \, \Omega_{c, i}}{2 \, \omega_{p, i}^2} \,
\frac{1}{\mu^2} \, \frac{ \biggl[ \biggl( 1 - \frac{1}{\mu}
\biggr)^2 + \frac{x^2}{\mu} \, \sin^2\theta \biggr]}{\biggl[
\biggl( 1 - \frac{1}{\mu} \biggr)^2 + \frac{x^2}{\mu}
\biggr]^{3/2}} \, . \label{Preduced}
\end{equation}

\section{Stability analysis \label{stab}}

The modulational stability profile of a carrier wave whose
amplitude is described by the NLS Equation (\ref{NLSE}) has long
been studied, so only the main results have to be summarized here
\cite{Newell, Remo, Fedele, IKPSDIAW, IKPSDAW}.

The analysis consists in considering the linear stability of the
monochromatic (Stokes's wave) solution of the NLSE (\ref{NLSE})
\(\psi \, = \, {\hat \psi} \, e^{i Q |\psi|^2 \tau} \, + \, c.c.
\).
If the product $P Q$ of the NLS coefficients is positive,
the wave's envelope may develop an instability when subject to an
external perturbation characterized by a wavenumber $\hat k$ lower than
$\hat k_{cr} = \sqrt{2 \frac{Q}{P}} |\hat\psi_{0}|$.
The instability growth rate \( \sigma =
|Im\hat\omega(\hat k)| \) then reaches its maximum
value for $\hat k = \hat k_{cr}/\sqrt{2}$, viz.
\( \sigma_{max} \,=\,
| Q |\,
|\hat\psi_{0}|^2
\).
On the other hand, the wave will be {\em stable} for all
values of $\hat k$ if the product $P  Q$ is {\em negative}.

\begin{figure}[htb]
 \centering
 \resizebox{3in}{!}{
 \includegraphics[]{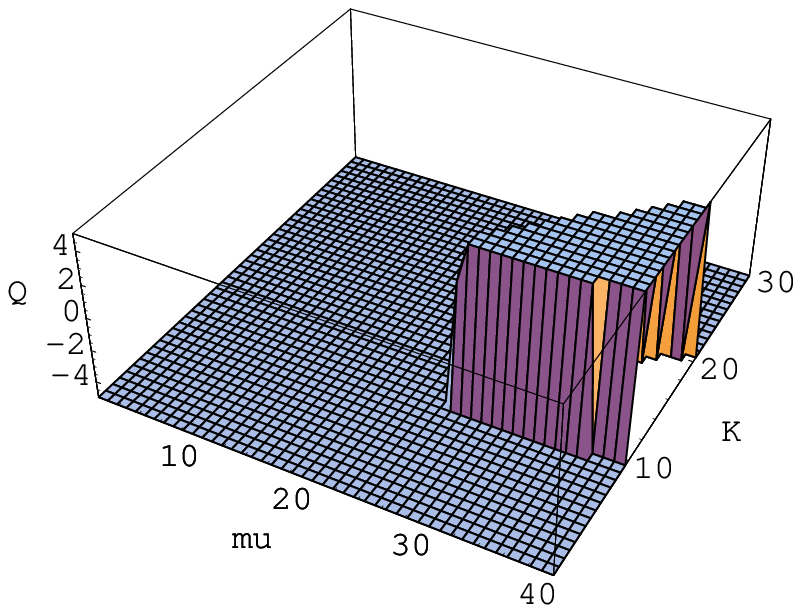}}
\caption{} \label{figure1}
\end{figure}

In our case, the dispersion coefficient $P$ is positive, so one
need only investigate the sign of the nonlinearity coefficient
$Q$, which is entirely determined by the quantity in brackets in
the right-hand-side of (\ref{Qreduced}); this is in fact a
bi-quadratic polynomial of $K$, say $p(K, \mu)$. It is a matter of
straightforward algebra to show (and an easy matter to confirm,
numerically) that $p(K, \mu)$ (and $Q$) is negative for values of
$\mu$ below $\mu_{cr} = 25.1146$, i.e. for all values of the
wavenumber $x$. Therefore, for negative dust charge ($s=-1$ i.e.
$\mu < 1$), the wave will always be stable. On the other hand, for
positive dust charge ($s=+1$ i.e. $\mu > 1$), the wave may become
unstable (only) for values of $\mu$ above $\mu_{cr}$ i.e. in the
case of positive dust charge concentration $q_d n_d$ higher than
$\approx 24 q_i n_i$ (a very rare situation, physically speaking,
which implies a very high ion depletion in the plasma). The
numerical value of $Q$, as expressed by relation (\ref{Qapprox}),
is roughly depicted in figure \ref{figure1} versus the wavenumber
$K$ and the dust parameter $\mu$. As predicted above, $Q$ (and $P
Q$) only reaches positive values for $\mu$ beyond $\approx 25$ and
$K$ above $\approx 10$ (i.e. $k > 10 \omega_{p, i}/c$), which is a
hardly ever realizable physical situation. We conclude that the
R-D-MHD waves are modulationally stable, in the presence of
negatively charged dust grains, and (practically)  also for
positively charged ones.

\section{Localized modes \label{localized}}

\begin{figure}[htb]
 \centering
 \resizebox{3in}{!}{
 \includegraphics[]{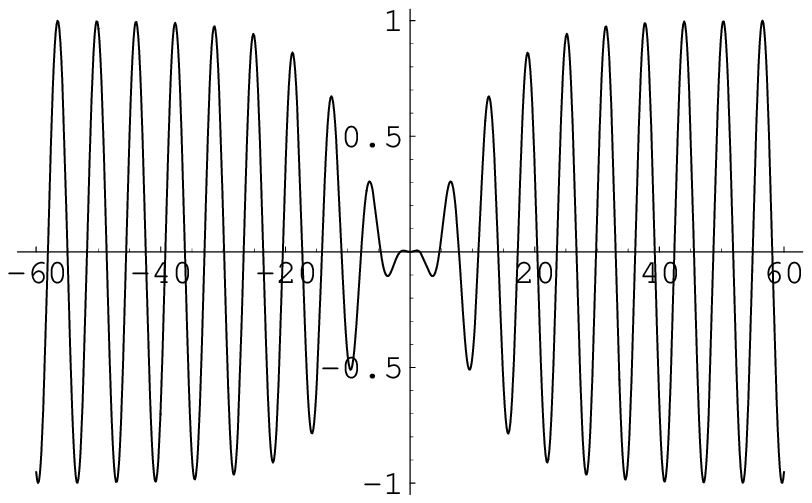}}
\\
\vskip 2 cm \resizebox{3in}{!}{
\includegraphics{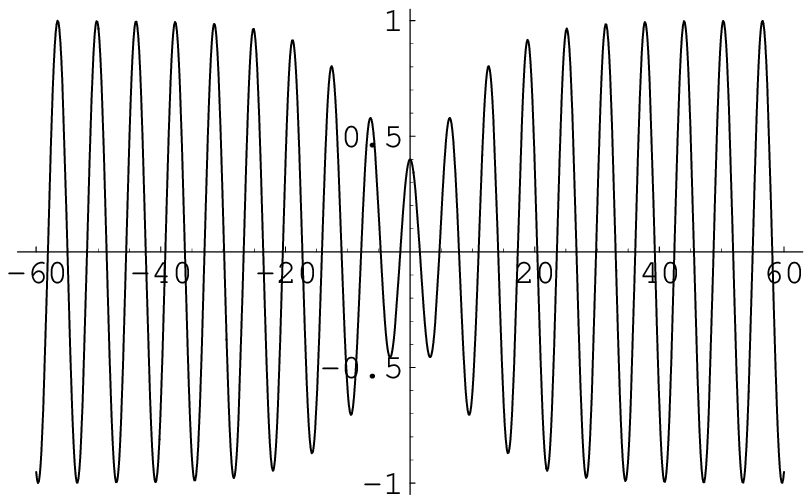}
} \caption{} \label{figure2}
\end{figure}

Different types of envelope excitations (solitons) are known to
satisfy Eq. (\ref{NLSE}); in specific, one finds bright- (dark- or
grey-) type solitons, e.g. pulses (holes) for a positive
(negative) value of the coefficient product $P Q$ \cite{Newell,
Remo, IKPSDIAW, IKPSDAW, Fedele}, as already long known from
nonlinear optics \cite{Newell2, Hasegawa}. According to the
conclusions of the preceding Section, the R-D-MHD waves considered
in this study will (in the majority of physically realizable
situations) rather favour dark-type localized excitations, i.e.
field dips (voids) propagating at a constant profile, thanks to
the balance between the wave dispersion and nonlinearity. The
analytical form of these excitations, depicted in Fig.
\ref{figure2}, reads $\psi(\zeta, \tau) = \sqrt{\rho(\zeta, \tau)}
\, e^{i\,\Theta(\zeta, \tau) }$, where
\begin{equation}
\rho = \rho_0 \,
\biggl[ 1 - a^2\, sech^2 \biggl(\frac{\zeta - u\, \tau}{L}
\biggr)\biggr] \, ;
\label{greysoliton}
\end{equation}
where $a$ is a real parameter measuring the depth of the field
void: $0 < a < 1$ ($a = 1$) corresponds to grey (black) solitons;
see Fig. \ref{figure2}a (\ref{figure2}b). The complex expressions
for the parameters $a$ and $\Theta$ in the above expression (as
well as related ones for bright solitons) can readily be found in
the references \cite{Newell, Remo, IKPSDIAW, IKPSDAW, Fedele,
Newell2, Hasegawa} and are omitted here. Note, however, that the
width $L$ of (both bright- and dark-types of) these localized
excitations depends on the maximum amplitude $\rho_0$ as \( L =
\sqrt{2 \bigl|{P}/({Q\,\rho_0})\bigr|} \); therefore, we retain
that for a given amplitude, the (absolute value of the)
coefficient ratio $P/Q$ expresses the square width of the soliton,
i.e. a pulse if $P Q > 0$ and a hole if $P Q < 0$. Inversely, for
a fixed width $L$, the quotient $P/Q$ expresses the amplitude
(height) of the solitary wave $\rho_0$.

In figures \ref{figure3} -- \ref{figure8}  we have depicted the
ratio $P/Q$ as expressed by the relations (\ref{Qreduced}),
(\ref{Preduced}), expressed in units, say: $P_0/Q_0 = \biggl(
\frac{c^2 \, \Omega_{c, i}}{2 \, \omega_{p, i}^2} \biggr)/\biggl(
\frac{Z_i^2 e^2} {m_i^2 \,c^2 \, \Omega_{c, i}} \biggr) = (m_i^2
c^4  \Omega_{c, i}^2)/(2 Z_i^2 e^2 \omega_{p, i}^2)$. In the
presence of negative dust ($\mu < 1$, see Fig. \ref{figure3}a),
the soliton width is seen to bear lower values (with a maximum for
higher $K$) as $\mu$ decreases; therefore, an increase in the
concentration of negative dust results in generally narrower
excitations, but with a peak at higher wavenumbers $K$. Also, for
a given $K$, the width is maximum for a certain value of $\mu$
(see Fig. \ref{figure3}b); the position of the maximum depends
only slightly on $\theta$ but rather strongly on $K$ (see Figs.
\ref{figure4}a, b). Finally, for a fixed value of $\mu$, the $P/Q$
vs. $K$ curve seems to have a maximum at $\theta = \pi/2$; see
Fig. \ref{figure5}: transverse modulation slightly favours higher
soliton widths. This maximum moves to higher $K$ with increasing
dust (i.e. decreasing $\mu$); cf. Figs. \ref{figure5}a,
\ref{figure5}b.

For positive dust ($\mu > 1$), see Figs. \ref{figure6} -- \ref{figure8},
we have similar qualitative results, yet generally lower values.
Once more, the angle variable does not seem to influence the soliton profile
dramatically.

\section{Conclusions}

In this paper, we have studied the 2D linear and nonlinear
propagation of R-D-MHD waves in a uniform cold magnetoplasma
composed of electrons, ions, and charged dust grains.  The
presence of immobile charged dust grains is responsible for the
ion rotation and a new cutoff frequency (non-existing in an
ordinary e--i plasma), which were reported by Rao in his classic
paper \cite{Rao1995}. The propagation of the modified dust
magnetoacoustic waves is possible due to the finite ion inertia
effect. The charged dust modifies the phase speed of the modified
magnetosonic waves. Furthermore, we have considered the amplitude
modulation of the R-D-MHD waves and have shown that
self-interactions among waves  result in the harmonic generation
and the amplitude modulation of a carrier R-D-MHD wave.  The wave
envelope has been shown to be stable against perturbations in a
wide range of physical parameter spaces. Finally, we have
discussed the possibility of localized envelope excitations
(mostly of the dark soliton type i.e.  localized field dips
propagating in the plasma) associated with the nonlinear R-D-MHD.

\begin{acknowledgments}
This work was supported by the European Commission (Brussels)
through the Human Potential Research and Training Network via the
project entitled: ``Complex Plasmas: The Science of Laboratory
Colloidal Plasmas and Mesospheric Charged Aerosols'' (Contract No.
HPRN-CT-2000-00140).
\end{acknowledgments}

\newpage

\begin{appendix}

\section{1st-order perturbation: Derivation of the
dispersion relation and 1st-harmonic
amplitudes}

Consider the system: (\ref{redeqne}) -- (\ref{last}), which describes
the evolution of $\mathbf{S} = (n_e, \,n_i,
\,\mathbf{u_e}, \, \mathbf{u_i}, \,\mathbf{E}, \,\mathbf{B})$.
By linearizing around the equilibrium state $\mathbf{S_0} = (n_{e,
0}, \,n_{i, 0}, \,\mathbf{0}, \, \mathbf{0}, \,\mathbf{0},
\,\mathbf{B_0})$ viz. $\mathbf{S} = \mathbf{S_0} + \mathbf{S_1}$
and assuming linear perturbations of the form: $\mathbf{S_1} =
\mathbf{\hat S_1}\, \exp i(\mathbf{k x} - \omega t) = \mathbf{\hat
S_1}\, \exp i(k x + k y - \omega t)$, we obtain a new system of (linear)
equations for the perturbation amplitudes ${(\hat S_1)}_j$:
\begin{equation}
- i \, \omega \, \hat n_{e, 1} + i \,\mathbf{k} \,(n_{e, 0} \,
\mathbf{\hat u_{e1}}) \, = 0 \, , \label{redeqne-lin}
\end{equation}
\begin{equation}
- i \omega \hat n_{i, 1} + i \, \mathbf{k} \, (n_{i, 0}
\,\mathbf{\hat u_{i1}}) \,= 0  \, , \label{redeqni-lin}
\end{equation}
\begin{equation}
\hat E_{1 x} = - \frac{1}{c}\, \hat u_{e1, y} \,B_0 \, ,
\label{Ex-lin}
\end{equation}
\begin{equation}
\hat E_{1 y} = \frac{1}{c}\, \hat u_{e1, x} \,B_0 \, ,
\label{Ey-lin}
\end{equation}
\begin{equation}
m_i \,\bigl( - i \, \omega \, \bigr) \hat u_{i1, x} \, = \, Z_i
\,e \,\biggl( \hat E_{1x} + \frac{1}{c} \hat u_{i, y} B_0 \biggr)
\, = \,+  \frac{Z_i e B_0}{c} \,\bigl( \hat u_{i, y} - \hat u_{e,
y} \bigr) \, , \label{redequix-lin}
\end{equation}
\begin{equation}
m_i \,\bigl( - i \, \omega \, \bigr) \hat u_{i1, y} \, = \, Z_i
\,e \,\biggl( \hat E_{1y} - \frac{1}{c} \hat u_{i, x} B_0 \biggr)
\, = \,-  \frac{Z_i e B_0}{c} \,\bigl( \hat u_{i, x} - \hat u_{e,
x} \bigr) \, , \label{redequiy-lin}
\end{equation}
\begin{equation}
i \, k_y \, \hat B_1 \, = \, \frac{4 \pi e}{c} \,\bigl(
Z_i n_{i, 0} \hat u_{i1, x} - n_{e, 0} \hat u_{e1, x} \bigr) \, ,
\label{dBdy-lin}
\end{equation}
\begin{equation}
i \, k_x \, \hat B_1 \, = \, - \frac{4 \pi e}{c} \,\bigl(
Z_i n_{i, 0} \hat u_{i1, y} - n_{e, 0} \hat u_{e1, y} \bigr) \, ,
\label{dBdx-lin}
\end{equation}
and
\begin{equation}
i \, k_x \, \hat  E_{1, y}  \, -
\, i \, k_y \, \hat  E_{1, x} \, =
\frac{1}{c}  \, (i\, \omega)  \, \hat B_1 \, ,
\label{last-lin}
\end{equation}
where only first harmonic terms were retained.
Now, eliminating the electron velocity amplitudes
from (\ref{redequix-lin}) -- (\ref{dBdx-lin})
(i.e. solving for $\hat u_{e1, j}$ in the latter two and
substituting in the former), one immediately obtains:
\begin{eqnarray}
i \, \omega \, \hat u_{i1, x} \, + s \Omega_{c, i} \frac{Z_d
n_{d,0}}{n_{e,0}} \, \hat u_{i1, y} \,& = & \, i \Omega_{c, i}
\frac{c}{4 \pi e n_{e,0}}\,
\hat B_{1}\, k_x \nonumber \\
i \, \omega \, \hat u_{i1, y} \, - s \Omega_{c, i} \frac{Z_d
n_{d,0}}{n_{e,0}} \, \hat u_{i1, x} \,& = & \, i \Omega_{c, i}
\frac{c}{4 \pi e n_{e,0}}\, \hat B_{1}\, k_y \label{u1xy-lin}
\end{eqnarray}
(the ion cyclotron frequency $\Omega_{c, i}$ was defined in the text).
Also, one may substitute from (\ref{Ex-lin}), (\ref{Ey-lin})
into (\ref{last-lin}) in order to obtain:
\begin{equation}
\mathbf{k \cdot \hat u_{e1}} = \frac{\omega}{B_{0}} \hat B_1 \, ,
\end{equation}
and, once more, use (\ref{redequix-lin}), (\ref{redequiy-lin})
to eliminate $\mathbf{u_{e1}}$ in it:
\begin{equation}
\frac{\omega}{B_{0}} \hat B_1 \, = \, \frac{Z_i n_{i, 0}}{n_{e,
0}} \, \mathbf{k \cdot \hat u_{i1}} \, . \label{lastlast-lin}
\end{equation}
Now, (\ref{u1xy-lin}a, b), (\ref{lastlast-lin}) form a closed system, with
respect to $\hat u_{i1, j}$ ($j = x, y$) and $\hat B_1$.
In specific, one may solve the latter for $\hat B_1$
and substitute into the former two; one thus obtains
precisely the system of equations (\ref{vxvy}), along with
the definitions mentioned in the text.

On a more systematic basis, one may define the matrix:
\begin{equation}
\mathbf{L}_0^{(l)} (\omega, \mathbf{k}) =
\left(
\begin{array}{ccccccccc}
- i l \omega & 0 & i l k_x n_{e, 0} & i l k_y n_{e, 0} & 0 & 0 & 0 & 0 & 0 \\
0 & - i l \omega & 0 & 0 & i l k_x n_{i, 0} & i l k_y n_{i, 0} &  0 & 0 & 0 \\
0 & 0 & 0 & B_0 & 0 & 0 &  c & 0 & 0  \\
0 & 0 & -B_0 & 0 & 0 & 0 &  0 & c & 0 \\
0 & 0 & 0 & \Omega_{c, i} & - i l \omega & - \Omega_{c, i} & 0 & 0 & 0 \\
0 & 0 & -\Omega_{c, i} & 0 & \Omega_{c, i} & - i l \omega & 0 & 0 & 0 \\
0 & 0 & \frac{4 \pi e}{c} n_{e, 0}  & 0 & -\frac{4 \pi e}{c} Z_i n_{i, 0}
& 0 & 0 & 0 & i l k_y \\
0 & 0 & 0 & -\frac{4 \pi e}{c}n_{e, 0} & 0 & \frac{4 \pi e}{c} Z_i n_{i, 0}
& 0  & 0 & i l k_x \\
0 & 0 & 0 & 0 & 0 & 0
& - i l k_y  & i l k_x & - i \frac{l \omega}{c}
\end{array}
\right) \, ,
\label{defL0}
\end{equation}
which arises naturally by isolating the $l-$th harmonic terms (at every order
$n$)
in equations (\ref{redeqne}) -- (\ref{last}).
For instance, for $n = l = 1$,
the system on top of this Appendix is formally expressed
as: $\mathbf{L}_0^{(1)} \, \mathbf{S}_1^{(l)} \, = \, \mathbf{0}$.
Now, the condition $Det\mathbf{L}_0^{(1)} = 0$ leads exactly to the dispersion
relation (\ref{dispersion}), while the solution of the system
is given by (\ref{ne1-solution}) -- (\ref{Ey1-solution}) in the text.

\section{2nd-order perturbation: group velocity, 0-th and 2-nd
harmonic amplitude corrections}

For $n = 2$, $l = 1$, we obtain the system of equations
$\mathbf{L}_0^{(1)} \, \mathbf{S}_1^{(2)} = \mathbf{R}_1^{(2)}$,
where $\mathbf{L}_0^{(1)}$ was defined in (\ref{defL0}) and
$\mathbf{R}_1^{(2)}$ denotes the vector:
$\bigl( \lambda c_1^{(11)} - n_{e, 0} c_3^{(11)}, \,
\lambda c_2^{(11)} - n_{i, 0} c_5^{(11)}, \,
0, \, 0, \,
\lambda c_5^{(11)}, \,
\lambda c_6^{(11)}, \,
0, \,
-1, \,
\frac{1}{c} \lambda - c_8^{(11)} \bigr)^T \, \partial B_1^{(1)} /\partial \zeta
$. The compatibility condition imposed in order for a solution
to exist, can be formulated as the constraint:
$Det\mathbf{L}_m^{(1)} = 0$, where $\mathbf{L}_m^{(1)}$ is the matrix
obtained by substituting the $m-$th column in $\mathbf{L}_0^{(1)}$ by
$\mathbf{R}_1^{(2)}$. Whichever the choice of $m$ ($= 1, 2, ..., 9$), by
solving the resulting equation, one readily obtains the definition of
$\lambda$ as the group velocity $v_g = \partial \omega/\partial k_x$
(as defined in the text).
One then obtains the solution
$S_{1, j}^{(2)} = c_{j}^{(21)} {\partial B_1^{(1)}}/{\partial
\zeta}$ for (8 of) the elements of $\mathbf{S_{2}}^{(1)}$, in terms of
one of them, e.g. of $S_{1, 9}^{(2)} = B_{1}^{(2)}$.
Assuming, with no loss of generality, that $B_{1}^{(2)} = 0$, one obtains
for the coefficients $c_{j}^{(21)}$ the expressions:
\begin{eqnarray}
c^{(21)}_1 \,&=& \, 0 \nonumber \\
c^{(21)}_2 \,&=& \, 0 \nonumber \\
c^{(21)}_3 \,&=& \, \frac{n_{i, 0} \, Z_i}{B_0^2 \, e \, n_{e, 0}^2 \,
n_{i, 0} \, Z_i^2 \,
\omega \, k^2}
\biggl\{
i \, B_0 \, e \, Z_i \biggl[ n_{e, 0}^2 (\omega^2 + \Omega_{c, i}^2)
- 2 n_{e, 0} n_{i, 0} Z_i \Omega_{c, i}^2 +
n_{i, 0}^2 \Omega_{c, i}^2 Z_i^2
\biggr] \cos^2\theta
\nonumber \\
& & \,
\qquad \qquad \qquad \qquad \qquad
%\qquad
+ \, n_{e, 0} \, \omega \, \Omega_{c, i}
\biggl[ - i \, c \, m_i \, n_{e, 0} \, \omega + B_0 e Z_i
(Z_i n_{i, 0} - n_{e, 0})
\sin2\theta
\biggr]
\biggr\}
\nonumber \\
c^{(21)}_4 \,&=& \, \frac{1}{2 \, B_0^2 \, e \, n_{e, 0}^2 \, n_{i, 0}
\, Z_i^2 \,
\omega \, k^2}
\biggl\{
2 \, c \, m_i \, n_{e, 0}^2 \, \omega \,
\bigl[ n_{e, 0} (\omega^2 - \Omega_{c, i}^2) +
n_{i, 0} \Omega_{c, i}^2 Z_i \bigr]
\nonumber \\
& & \,
\qquad \qquad \qquad \qquad \qquad +
B_0 \, e \, n_{i, 0} \, Z_i^2
\biggl[ 4 n_{e, 0} \, \omega \, \Omega_{c, i}
(n_{e, 0} - Z_i n_{i, 0}) \cos^2\theta
\nonumber \\
& & \qquad \qquad \qquad \qquad \qquad \qquad \,
i \,
\bigl[ n_{e, 0}^2 (\omega^2 + \Omega_{c, i}^2)
- 2 n_{e, 0} n_{i, 0} \Omega_{c, i}^2 Z_i +
n_{i, 0}^2  \Omega_{c, i}^2 Z_i^2
\bigr]
\sin2\theta
\biggr]
\biggr\}
\nonumber \\
c^{(21)}_5 \,&=& \, \frac{i}{8 \, \pi\, m_i^2\, n_{e, 0} \, n_{i, 0}
\,
\omega \,c^2\, k^2}
\biggl\{
B_0 \, Z_i \, \bigl[ m_i \, n_{i, 0}
\,
c^2\, k^2 + 8 \, \pi\, e^2\,(n_{e, 0} - Z_i n_{i, 0})^2 \bigr]
\cos2\theta \nonumber \\
& & \qquad \qquad \qquad \qquad \qquad \qquad
+ c\, m_i\,
\bigl[ - B_0 \, Z_i n_{i, 0}\, c \, k^2 +
8 \pi i e \, n_{e, 0}\, \omega \, (n_{e, 0} - Z_i n_{i, 0})
\sin2\theta
\biggr\}
\nonumber \\
c^{(21)}_6 \,&=& \, \frac{1}{8 \, \pi\, m_i^2\, n_{e, 0} \, n_{i, 0}
\,
\omega \,c^2\, k^2}
\biggl\{
i \, B_0 \, Z_i \, \bigl[ m_i \, n_{i, 0}
\,
c^2\, k^2 + 8 \, \pi\, e^2\,(n_{e, 0} - Z_i n_{i, 0})^2 \bigr]
\sin2\theta \nonumber \\
& & \qquad \qquad \qquad \qquad \qquad \qquad
+
8 \pi \,e \, c\, m_i\, n_{e, 0}\, \omega \, (n_{e, 0} - Z_i n_{i, 0})
\cos2\theta
\biggr\} \nonumber \\
c^{(21)}_7 \,&=& \,\frac{B_0}{8 \, \pi\, e m_i^2\, n_{e, 0}^2 \,
\omega \,c^3\, k^2} \biggl\{ - 2 \, m_i^2 \, n_{e, 0} \, \omega \,
c^3\, k^2 + \, e\,Z_i \biggl[ - 8 \pi e \, c \, m_i \, n_{e, 0}
\omega (n_{e, 0} - Z_i n_{i, 0}) \cos2\theta
 \nonumber \\
& & \qquad \qquad \qquad \qquad \qquad \qquad - i B_0 \, Z_i \,
\bigl[ m_i n_{i, 0}\, c^2 \, k^2 + 8 \pi e^2 \, (n_{e, 0} - Z_i
n_{i, 0})^2 \bigr] \sin2\theta \biggr]
\biggr\} \nonumber \\
c^{(21)}_8 \,&=& \, \frac{i B_0 \, Z_i }{8 \, \pi\, e m_i^2\,
n_{e, 0}^2 \, \omega \,c^3\, k^2} \biggl\{c\,m_i \biggl[ - B_0  \,
c \,k^2 n_{i, 0}\, Z_i \, + 8 \pi i e n_{e, 0}\,
\omega (n_{e, 0} - Z_i n_{i, 0}) \sin 2\theta \nonumber \\
& & \qquad \qquad \qquad \qquad \qquad  \qquad+ B_0 \, Z_i \,
\bigl[ m_i n_{i, 0}\, c^2 \, k^2 + 8 \pi e^2 \, (n_{e, 0} - Z_i
n_{i, 0})^2 \bigr] \cos 2\theta \biggr]
\biggr\} \nonumber \\
c^{(21)}_9 \,&=&\, 0
\label{App-c21j}
\end{eqnarray}

For $n = 2$, $l = 0$, we obtain the system of equations
$\mathbf{L}_0^{(0)} \, \mathbf{S}_0^{(2)} = \mathbf{R}_0^{(2)} \,
|B_1^{(1)}|^2$ (set $l = 0$ in (\ref{defL0}) for
$\mathbf{L}_0^{(0)}$), where $\mathbf{R}_0^{(2)}$ is the vector
$\bigl( 0, \, 0, \, -c_4^{(11)}, \, c_3^{(11)}, \, i k_y
c_6^{(11)} {c_5^{(11)}}^* + \frac{\Omega_{c, i}}{B_0} (c_6^{(11)}
- c_4^{(11)}), \, - i k_x c_6^{(11)} {c_5^{(11)}}^* -
\frac{\Omega_{c, i}}{B_0} (c_5^{(11)} - c_3^{(11)}), \, \frac{4
\pi e}{c} (Z_i c_2^{(11)} {c_5^{(11)}}^* - c_1^{(11)}
{c_3^{(11)}}^*) , \, - \frac{4 \pi e}{c} (Z_i c_2^{(11)}
{c_6^{(11)}}^* - c_1^{(11)} {c_4^{(11)}}^*), \, 0
  \bigr)^T    + c.c.$.
 The 1st, 2nd and 9th equations are identically satisfied, so
 the corresponding equations for $n = 3, l = 0$ have to be
 ``borrowed''. Combining them with
 the remaining (3rd to 8th) equations here, we obtain
 \begin{eqnarray}
 {n_{e}}_0^{(2)} \, &=& \, \frac{n_{e, 0}} {v_g}
c^{(20)}_3 \,|B_1^{(1)}|^2
 \, ,
 \nonumber \\
 {n_{i}}_0^{(2)} \, &=& \, \frac{n_{i, 0}} {v_g}
c^{(20)}_5 \,|B_1^{(1)}|^2
 \, ,
 \nonumber \\
{u_{e, x}}_0^{(2)} \, &=& \, - \frac{2 \omega} {B_0^2 k} \cos\theta
\,|B_1^{(1)}|^2 \equiv
c^{(20)}_3 \,|B_1^{(1)}|^2
 \, ,
 \nonumber \\
 {u_{e, y}}_0^{(2)} \, &=& \, - \frac{2 \omega} {B_0^2 k} \sin\theta
\,|B_1^{(1)}|^2 \equiv
c^{(20)}_4 \,|B_1^{(1)}|^2
 \, ,
 \nonumber \\
 {u_{i, x/y}}_0^{(2)} \, &=& \, \biggl( \frac{n_{e, 0}} {Z_i n_{i, 0}}  \biggr)^2
 {u_{e, x/y}}_0^{(2)} \equiv
c^{(20)}_{5/6} \,|B_1^{(1)}|^2
 \, ,
 \nonumber \\
  {E_{x}}_0^{(2)} \, &=& \,  {E_{y}}_0^{(2)} \, = \,  {B}_0^{(2)} \, = \,  0
  \label{App-c20j}
\end{eqnarray}

For $n = 2$, $l = 2$, we obtain a system of (9) equations in the
matrix form: $\mathbf{L}_0^{(2)} \, \mathbf{S}_2^{(2)} =
\mathbf{R}_2^{(2)}\, {B_1^{(1)}}^2$ [set $l = 2$ in (\ref{defL0}]
for $\mathbf{L}_0^{(2)}$); the (lengthy) expression of the vector
$\mathbf{R}_2^{(2)}$ is omitted. Solving for the second-harmonic
amplitudes ${S_{j}}_2^{(2)}$, we obtain:
 \begin{eqnarray}
 {n_{e}}_2^{(2)} \, &=& \, \frac{n_{e, 0}^2 \, \bigl[ m_i n_{i, 0} c^2 \, k^2 + 4 \pi e^2
 (n_{e, 0} - n_{i, 0} Z_i)^2 \bigr]}{2 \pi e^2 n_{i, 0} Z_i B_0^2
 (n_{e, 0} - n_{i, 0} Z_i)^2} \,{B_1^{(1)}}^2 \equiv
c^{(22)}_1 \,{B_1^{(1)}}^2
 \, ,
 \nonumber \\
 {n_{i}}_2^{(2)} \, &=& \, \frac{1}{Z_i} c^{(22)}_1 \,{B_1^{(1)}}^2 \equiv
c^{(22)}_2 \,{B_1^{(1)}}^2
 \, ,
 \nonumber \\
{u_{e, x}}_2^{(2)} \, &=& \, \frac{c^3 m_i^3 \, \omega}{3\pi e^4
\, n_{i, 0} \,
 Z_i^4 \, B_0^5 (n_{e, 0} - n_{i, 0} Z_i)^2 \, k}\, \times  \nonumber
 \\ &&
 \biggl\{
 \Omega_{c, i}\,
 \biggl[ 6 \pi \, e \, n_{e, 0}^3\, \omega^2 + \pi \, Z_i
 \, e \,  n_{e, 0}^2\,n_{i, 0} \,
 (- 4 \omega^2 +  \Omega_{c, i}^2)
 \nonumber \\ &&
\qquad \qquad  + n_{i, 0}^2 \,\Omega_{c, i}\, Z_i^2 (B_0 \, c\,
k^2 - 2 \pi e n_{e, 0} \, \Omega_{c, i} ) + \pi e Z_i^3 \, \,
n_{i, 0}^3 \,\Omega_{c, i}^2 \biggr] \, \cos\theta \nonumber \\ &
& - i\, n_{e, 0}\, \omega \, \biggl[ B_0 \, c\, k^2 \,\Omega_{c,
i}\, (3 n_{e, 0}\,+ Z_i n_{i, 0})+ \pi \, e \, \bigl[ - 2 \,n_{e,
0}^2 (2 \omega^2 + \Omega_{c, i}^2) \nonumber \\ & & \qquad \qquad
+ \,n_{e, 0}\,n_{i, 0}\, Z_i \Omega_{c, i}^2 + n_{i, 0}^2 \,
Z_i^2\, \Omega_{c, i}^2 \bigr] \biggr]\, \sin\theta
\biggr\}
\nonumber \\ & &
 \equiv c^{(22)}_3 \,{B_1^{(1)}}^2
 \nonumber \\
 {u_{e, y}}_2^{(2)} \, &=& \,  \frac{c^3 m_i^3 \, \omega}{3\pi e^4
\, n_{i, 0} \,
 Z_i^4 \, B_0^5 (n_{e, 0} - n_{i, 0} Z_i)^2  \, k}\, \times  \nonumber
 \\ &&
 \biggl\{
 \Omega_{c, i}\,
 \biggl[ 6 \pi \, e \, n_{e, 0}^3\, \omega^2 + \pi \, Z_i
 \, e \,  n_{e, 0}^2\,n_{i, 0} \,
 (- 4 \omega^2 +  \Omega_{c, i}^2)
 \nonumber \\ &&
\qquad \qquad  + n_{i, 0}^2 \,\Omega_{c, i}\, Z_i^2 (B_0 \, c\,
k^2 - 2 \pi e n_{e, 0} \, \Omega_{c, i} ) + \pi e Z_i^3 \, \,
n_{i, 0}^3 \,\Omega_{c, i}^2 \biggr] \, \sin\theta \nonumber \\ &
& + \,  i\, n_{e, 0}\, \omega \, \biggl[ B_0 \, c\, k^2
\,\Omega_{c, i}\, (3 n_{e, 0}\,+ Z_i n_{i, 0})+ \pi \, e \, \bigl[
- 2 \,n_{e, 0}^2 (2 \omega^2 + \Omega_{c, i}^2) \nonumber \\ & &
\qquad \qquad + \,n_{e, 0}\,n_{i, 0}\, Z_i \Omega_{c, i}^2 + n_{i,
0}^2 \, Z_i^2\, \Omega_{c, i}^2 \bigr] \biggr]\, \cos\theta
\biggr\}  \nonumber \\ & &
 \equiv
c^{(22)}_4 \,{B_1^{(1)}}^2
 \, ,
 \nonumber \\
 {u_{i, x}}_2^{(2)} \, &=& \, \frac{c^2 m_i^2 \, n_{e, 0}^2 \, \omega}
 {3\pi e^3
\, n_{i, 0}^2 \,
 Z_i^4 \, B_0^4 (n_{e, 0} - n_{i, 0} Z_i)^2 \, k}\, \times  \nonumber
 \\ &&
 \biggl\{
  \biggl[
  B_0 \, c\, k^2 \, Z_i n_{i, 0}\, \omega + \pi \, e\,
  [n_{e, 0}^2 (2 \omega^2 +  \Omega_{c, i}^2) - 2 n_{e, 0}\, n_{i,
  0}\,Z_i \Omega_{c, i}^2 \, + Z_i^2 n_{i, 0}^2 \Omega_{c, i}^2 ]
  \biggr]
  \cos\theta
\nonumber \\
&& \qquad \qquad + 3  \pi \, i \,e \, n_{e, 0}\,\Omega_{c, i} \,
\omega \, (n_{e, 0} - n_{i, 0} Z_i) \, \sin\theta \biggr\}
  \equiv
c^{(22)}_5 \,{B_1^{(1)}}^2
 \, ,
 \nonumber \\
  {u_{i, y}}_2^{(2)} \,&=& \,  \frac{c^2 m_i^2 \, n_{e, 0}^2 \, \omega}
 {3\pi e^3
\, n_{i, 0}^2 \,
 Z_i^4 \, B_0^4 (n_{e, 0} - n_{i, 0} Z_i)^2 \, k}\, \times  \nonumber
 \\ &&
 \biggl\{
  \biggl[
  B_0 \, c\, k^2 \, Z_i n_{i, 0}\, \omega + \pi \, e\,
  [n_{e, 0}^2 (2 \omega^2 +  \Omega_{c, i}^2) - 2 n_{e, 0}\, n_{i,
  0}\,Z_i \Omega_{c, i}^2 \, + Z_i^2 n_{i, 0}^2 \Omega_{c, i}^2 ]
  \biggr]
  \sin\theta
\nonumber \\
&& \qquad \qquad - 3  \pi \, i \,e \, n_{e, 0}\,\Omega_{c, i} \,
\omega \, (n_{e, 0} - n_{i, 0} Z_i) \, \cos \theta \biggr\} \equiv
c^{(22)}_6 \,{B_1^{(1)}}^2
 \, ,
 \nonumber \\
  {E_{x}}_2^{(2)} \, &=&\,    - \frac{m_i^2 \, c}
 {3 \pi e^3
\, n_{i, 0} \,
 Z_i^3 \, B_0^3 (n_{e, 0} - n_{i, 0} Z_i)^2 \, k}\, \times  \nonumber
 \\ &&
 \biggl\{
 i \biggl[
 B_0 c\, k^2 \, (3 n_{e, 0}^2 \omega^2 + n_{e, 0} n_{i, 0} Z_i \Omega_{c, i}^2 -
 n_{i, 0}^2 Z_i^2 \Omega_{c, i}^2) +
\nonumber
 \\ &&
 \pi e \Omega_{c, i}\,(n_{e, 0} - n_{i, 0} Z_i)
 [ n_{e, 0}^2 \, (-7 \omega^2 + \Omega_{c, i}^2) - 2 \,
 n_{e, 0} n_{i, 0} Z_i\,\Omega_{c, i}^2 \,+ n_{i, 0}^2 Z_i^2\,\Omega_{c, i}^2
]\biggr] \cos\theta \nonumber
 \\ &&
+ 6 \pi \, e \, n_{e, 0}^3\, \omega^3 \, \sin\theta \biggr\}
 \equiv
c^{(22)}_7 \,{B_1^{(1)}}^2
 \nonumber \\
  {E_{y}}_2^{(2)} \, &=& \,  \frac{m_i^2 \, c}
 {3 \pi e^3
\, n_{i, 0} \,
 Z_i^3 \, B_0^3 (n_{e, 0} - n_{i, 0} Z_i)^2 \, k}\, \times  \nonumber
 \\ &&
 \biggl\{ -
 i \biggl[
 B_0 c\, k^2 \, (3 n_{e, 0}^2 \omega^2 + n_{e, 0} n_{i, 0} Z_i \Omega_{c, i}^2 -
 n_{i, 0}^2 Z_i^2 \Omega_{c, i}^2) +
\nonumber
 \\ &&
 \pi e \Omega_{c, i}\,(n_{e, 0} - n_{i, 0} Z_i)
 [ n_{e, 0}^2 \, (-7 \omega^2 + \Omega_{c, i}^2) - 2 \,
 n_{e, 0} n_{i, 0} Z_i\,\Omega_{c, i}^2 \,+ n_{i, 0}^2 Z_i^2\,\Omega_{c, i}^2
]\biggr] \sin\theta \nonumber
 \\ &&
+ 6 \pi \, e \, n_{e, 0}^3\, \omega^3 \, \cos\theta \biggr\}
\equiv c^{(22)}_8 \,{B_1^{(1)}}^2
 \nonumber \\
  {B}_2^{(2)} \, &=& \,   \frac{2 c^2 m_i^2 \, n_{e, 0}^3 \, \omega^2}
 {e^2
\, n_{i, 0} \,
 Z_i^3 \, B_0^3 (n_{e, 0} - n_{i, 0} Z_i)^2} \equiv
c^{(22)}_9 \,{B_1^{(1)}}^2 \, .
  \label{App-c22j}
\end{eqnarray}

\end{appendix}

\newpage

% FIGURE CAPTIONS

\centerline{\textbf{Figure Captions}}

Figure 1.

The value of the coefficient $Q$ is depicted against the dust parameter $\mu$ and
the (normalized) wavenumber $K$.

\vskip 1cm

Figure 2.

Soliton solutions of the NLS equation for $P Q < 0$ (holes); these
excitations are of: (a) dark type, (b) grey type. Notice that the
amplitude never reaches zero in (b). These excitations represent
electromagnetic field dips (voids) associated with the nonlinear
R-D-MHD wave propagation.

\vskip 1cm

Figure 3.

Negative dust; the (normalized) soliton width $L$ (absolute value
of $P/Q$) is depicted: (a) against wavenumber $K$, for $\theta =
0$ and $\mu = 0.8, 0.7, 0.6, 0.5$ (from top to bottom); (b)
against the dust parameter $\mu$, for $K = 0.2$ and $\theta =
0{}^\circ, 30{}^\circ, 60{}^\circ, 90{}^\circ$ (from bottom to
top).

\vskip 1cm

Figure 4.

Negative dust; the (normalized) soliton width  $L$ (absolute value
of $P/Q$) is depicted versus the dust parameter $\mu$ and the
angle $\theta$ for: (a) $K = 0.2$; (b) $K = 0.5$.

\vskip 1cm

Figure 5.

Negative dust; the soliton width  $L$ (absolute value of $P/Q$) is
depicted versus the wavenumber $K$ and the angle $\theta$ for: (a)
$\mu = 0.8$; (b) $\mu = 0.5$.

\vskip 1cm

Figure 6.

\begin{figure}[htb]
 \centering
 \resizebox{3in}{!}{
 \includegraphics[]{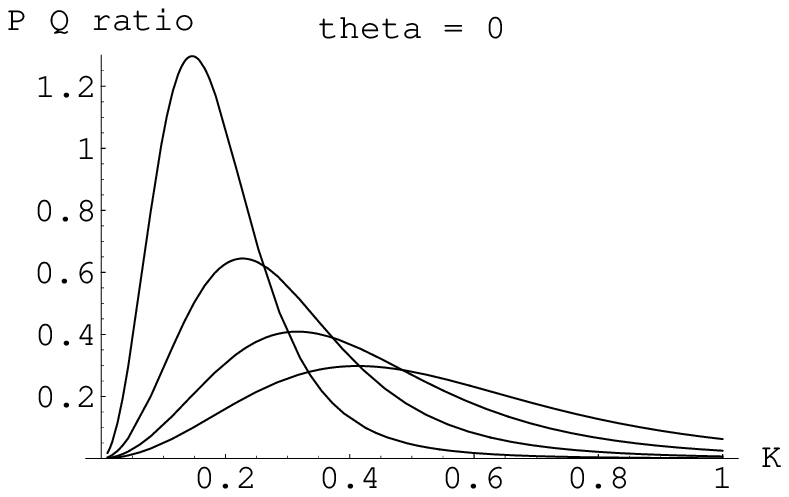}}
\\
\vskip 2 cm \resizebox{3in}{!}{
\includegraphics{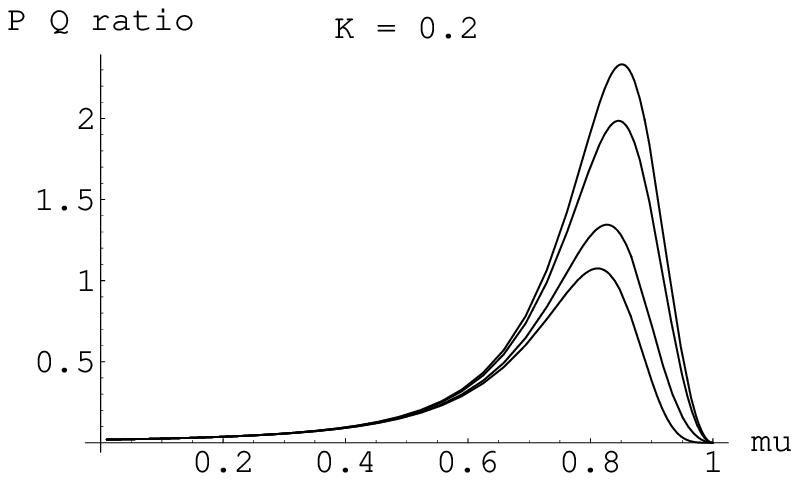}
} \caption{} \label{figure3}
\end{figure}

Similar to Fig. \ref{figure3}, for positive dust; the (normalized)
soliton width (absolute value of $P/Q$) is depicted: (a) against
wavenumber $K$, for $\theta = 0$ and $\mu = 1.5, 2.0, 2.5$ (from
top to bottom); (b) against the dust parameter $\mu$, for $K =
0.2$ and $\theta = 0{}^\circ, 30{}^\circ, 60{}^\circ, 90{}^\circ$
(from bottom to top).

\vskip 1cm

Figure 7.

\begin{figure}[htb]
 \centering
 \resizebox{3in}{!}{
 \includegraphics[]{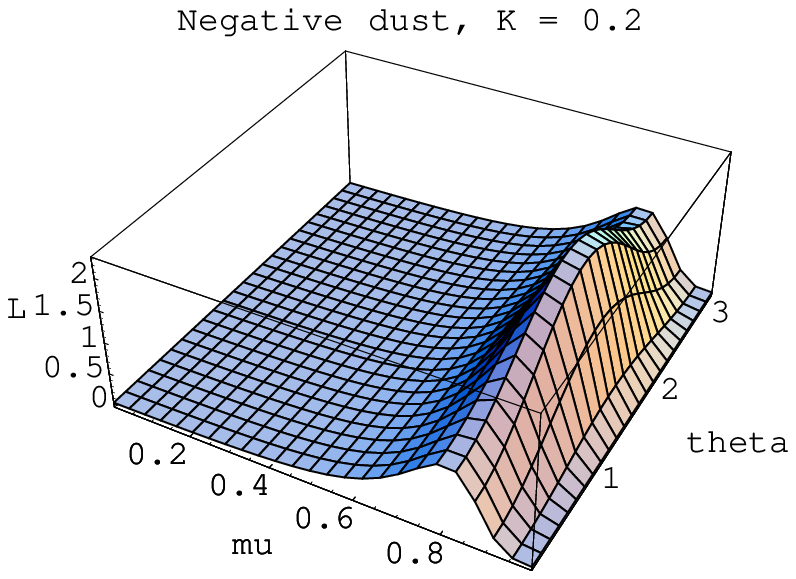}}
\\
\vskip 2 cm \resizebox{3in}{!}{
\includegraphics{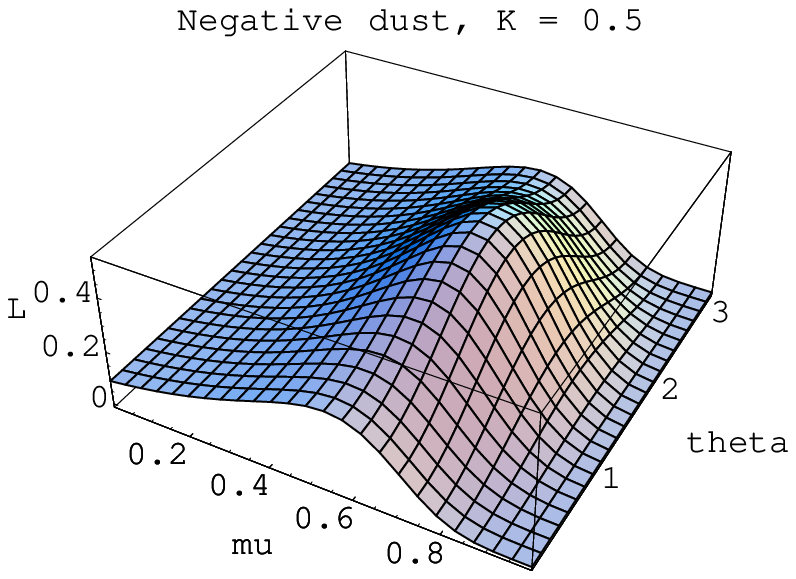}
} \caption{} \label{figure4}
\end{figure}

Similar to Fig. \ref{figure4}, for positive dust; the (normalized)
soliton width (absolute value of $P/Q$) is depicted versus the
dust parameter $\mu$ and the angle $\theta$ for: (a) $K = 0.2$;
(a) $K = 0.5$.

\vskip 1cm

Figure 8.

\begin{figure}[htb]
 \centering
 \resizebox{3in}{!}{
 \includegraphics[]{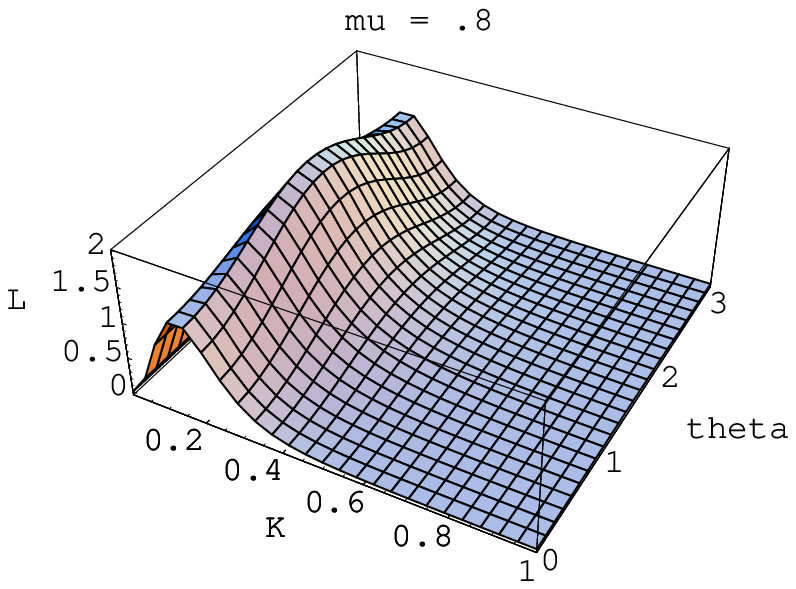}}
\\
\vskip 2 cm \resizebox{3in}{!}{
\includegraphics{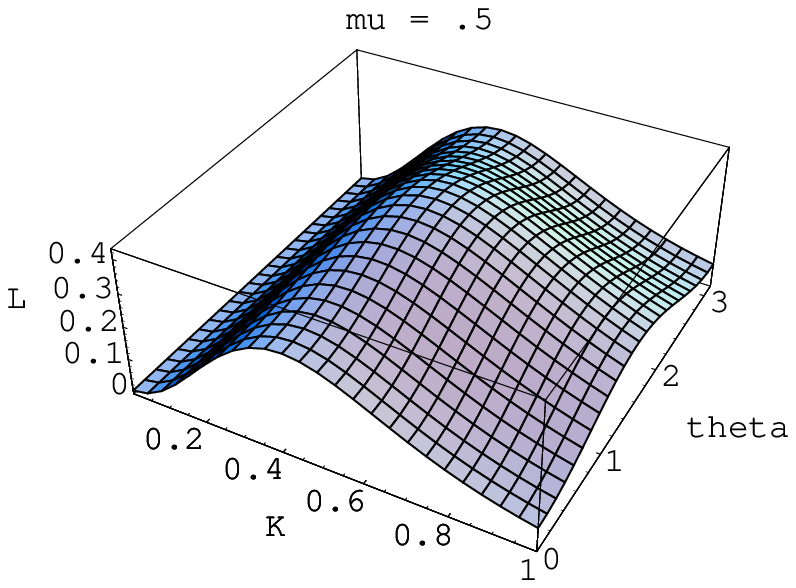}
} \caption{} \label{figure5}
\end{figure}

\begin{figure}[htb]
 \centering
 \resizebox{3in}{!}{
 \includegraphics[]{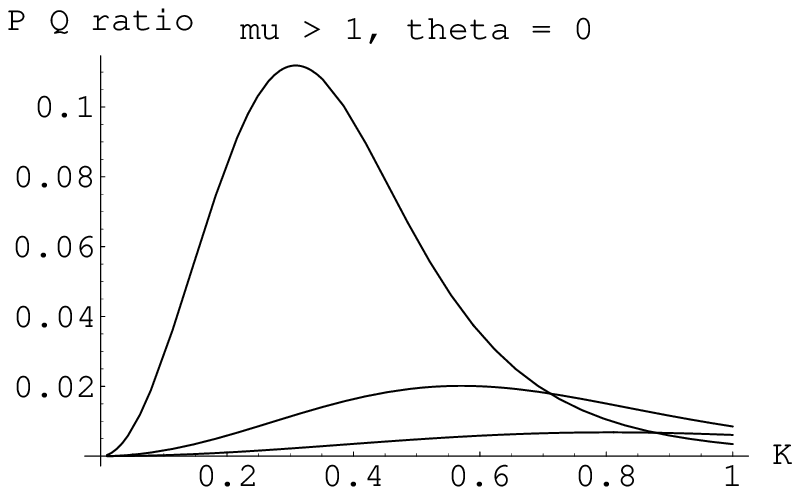}}
\\
\vskip 2 cm \resizebox{3in}{!}{
\includegraphics{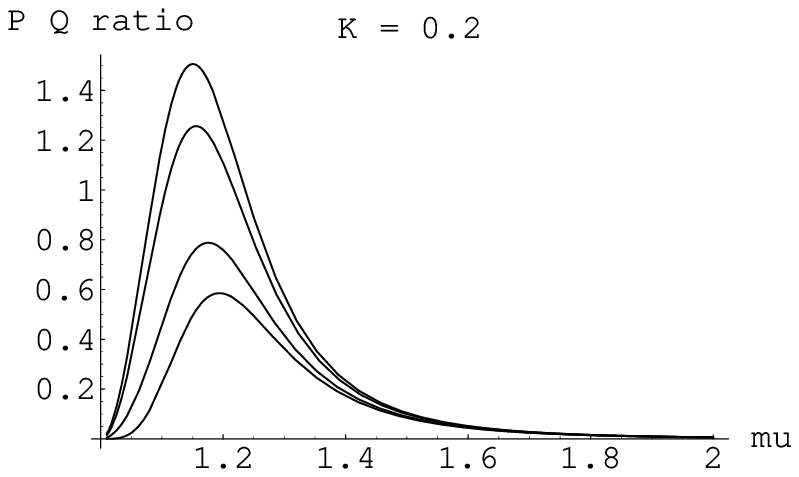}
} \caption{} \label{figure6}
\end{figure}

\begin{figure}[htb]
 \centering
 \resizebox{3in}{!}{
 \includegraphics[]{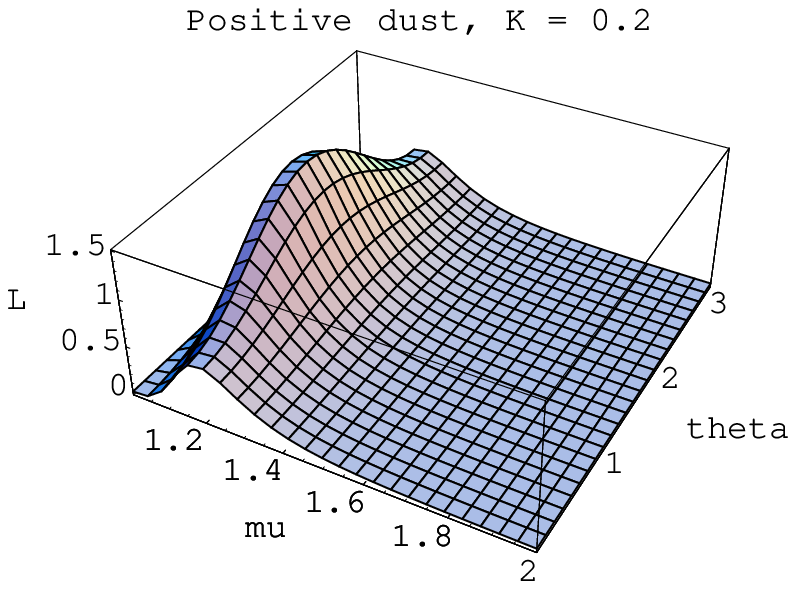}}
\\
\vskip 2 cm \resizebox{3in}{!}{
\includegraphics{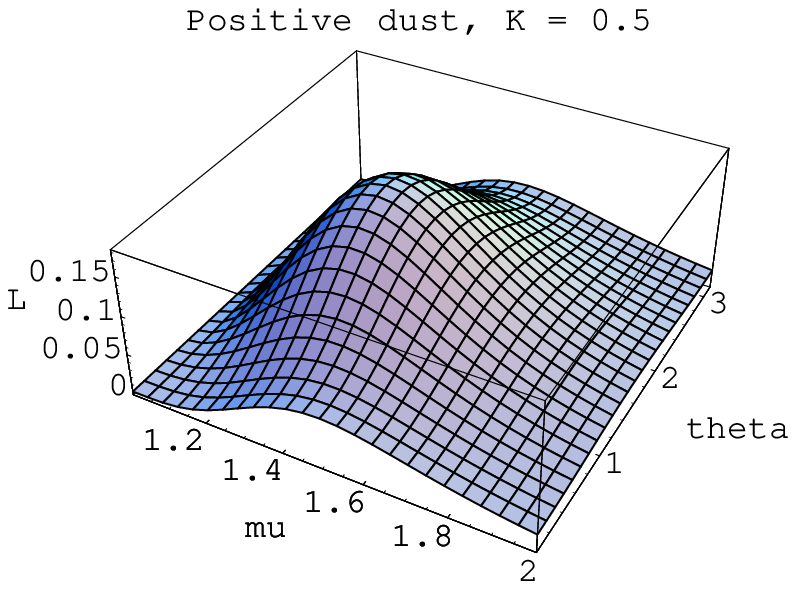}
} \caption{} \label{figure7}
\end{figure}

\begin{figure}[htb]
 \centering
 \resizebox{3in}{!}{
 \includegraphics[]{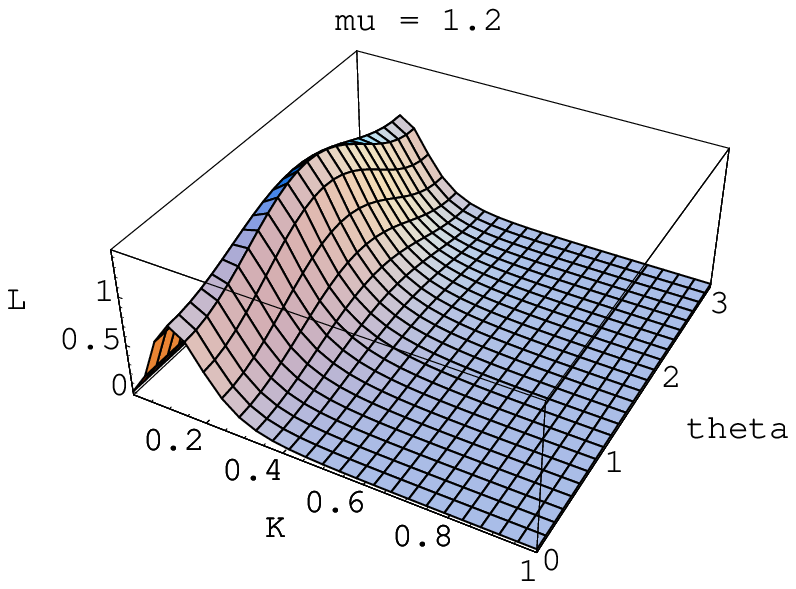}}
\\
\vskip 2 cm \resizebox{3in}{!}{
\includegraphics{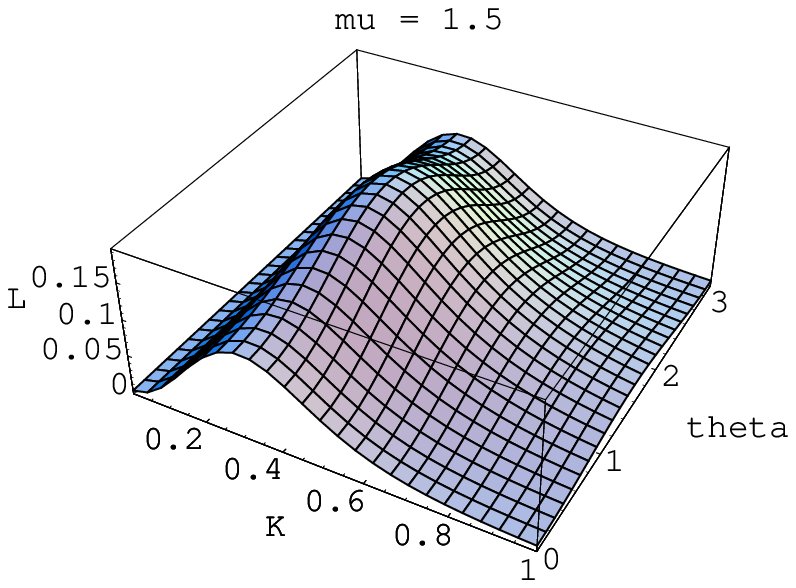}
} \caption{} \label{figure8}
\end{figure}

Similar to Fig. \ref{figure5}, for positive dust; the soliton
width (absolute value of $P/Q$) is depicted versus the wavenumber
$K$ and the angle $\theta$ for: (a) $\mu = 1.2$; (b) $\mu = 1.5$.

\newpage

\vskip 2 cm

% figure 1

\newpage

\vskip 2 cm

% figure 2

\newpage

% figure 3

\newpage

\newpage

% figure 4

\newpage

% figure 5
\newpage

% figure 6

\newpage

% figure 7

\newpage

% figure 8

\end{document}